\documentclass[aps,prl,reprint,superscriptaddress]{revtex4-2}

\pdfoutput=1

\usepackage{graphics,epsfig}
\usepackage{amsmath,amssymb}
\usepackage{bm,color}

\begin{document}
\title{Dynamical Crystallites of Active Chiral Particles}
\author{Zhi-Feng Huang}
\affiliation{Department of Physics and Astronomy, Wayne State University,
  Detroit, Michigan 48201, USA}
\author{Andreas M. Menzel}
\affiliation{Institut f\"{u}r Theoretische Physik II,
  Heinrich-Heine-Universit\"{a}t D\"{u}sseldorf, Universit\"atsstra{\ss}e 1, D-40225 D\"{u}sseldorf, Germany}
\affiliation{Institut f\"ur Physik, Otto-von-Guericke-Universit\"{a}t Magdeburg,
  Universit\"atsplatz 2, 39114 Magdeburg, Germany}
\author{Hartmut L\"{o}wen}
\affiliation{Institut f\"{u}r Theoretische Physik II,
  Heinrich-Heine-Universit\"{a}t D\"{u}sseldorf, Universit\"atsstra{\ss}e 1,  D-40225 D\"{u}sseldorf, Germany}

\begin{abstract}
One of the intrinsic characteristics of far-from-equilibrium systems is the
nonrelaxational nature of the system dynamics, which leads to novel properties
that cannot be understood and described by conventional pathways based on
thermodynamic potentials. Of particular interest are the formation and evolution
of ordered patterns composed of active particles that exhibit collective behavior.
Here we examine such a type of nonpotential active system, focusing on effects of
coupling and competition between chiral particle self-propulsion and self-spinning.
It leads to the transition between three bulk dynamical regimes dominated by
collective translative motion, spinning-induced structural arrest, and dynamical
frustration. In addition, a persistently dynamical state of self-rotating crystallites
is identified as a result of a localized-delocalized transition induced by the
crystal-melt interface. The mechanism for the breaking of localized bulk states can
also be utilized to achieve self-shearing or self-flow of active crystalline layers.
\end{abstract}

\date{\today}
\maketitle

Systems of self-propelled or self-spinning active particles are intrinsically out
of equilibrium. Operating with self-sustaining energetic sources or propulsive forces,
the corresponding active dynamic processes should no longer be governed by the
traditional relaxational pathways directed by the minimization principle of
thermodynamic potentials as in near-equilibrium samples of passive particles.
Instead, the system evolution is controlled by nonpotential, nonrelaxational dynamics,
a category that has been known to exhibit a variety of complex states such as
evolving ordered and defected patterns \cite{CrossRMP93,CrossPRL95,HuangPRE07},
spatiotemporal chaos with persistent dynamics \cite{MorrisPRL93,XiPRL93}, or glassy
behavior \cite{CugliandoloPRL97} as found in physical and biochemical pattern-forming
systems like fluid convection, liquid crystals, chemical reactions, and many biological
processes \cite{CrossRMP93}. Complex dynamical behavior has also been observed in
active colloidal materials both experimentally and computationally, varying from
phase separation \cite{Schwarz-LinekPNAS12,LiebchenPRL17,LevisPRE19,LeiSciAdv19,
SinghPRL19,CapriniPRL20}, dynamical clustering
\cite{Schwarz-LinekPNAS12,ButtinoniPRL13,KaraniPRL19,NiuACSNano18,vanderLindenPRL19},
active glass \cite{BerthierPRL14,KlongvessaPRL19},
to traveling \cite{MenzelPRL13,MenzelPRE14,PraetoriusPRE18,OphausPRE18}
or rotating \cite{NguyenPRL14,vanZuidenPNAS16,LiuPNAS20} crystals.

Among these dynamical phenomena of active matter, a common feature is the
collective motion of the constituent particles or building blocks in homogeneous
\cite{vanZuidenPNAS16,DasbiswasPNAS18,TsaiPRL05,YangPRE20,CapriniSoftMatter19,LiuPNAS20}
or phase-separated \cite{Schwarz-LinekPNAS12,LiebchenPRL17,LevisPRE19,SinghPRL19,
CapriniPRL20,LeiSciAdv19} liquid/gas states and ordered phases with different
crystalline symmetry \cite{MenzelPRL13,MenzelPRE14,PraetoriusPRE18,OphausPRE18,NguyenPRL14,
vanZuidenPNAS16,LiuPNAS20,SnezhkoCOCIS16,HanSciAdv20,ShenSoftMatter19,YeoPRL15,YangPRE17}.
In many cases the self-propulsion of active particles, coupled with interparticle
interactions, is the driving factor underlying various forms of collective behavior
\cite{Schwarz-LinekPNAS12,ButtinoniPRL13,MenzelPRL13,MenzelPRE14,PraetoriusPRE18,
OphausPRE18,vanderLindenPRL19,SinghPRL19,CapriniPRL20,BerthierPRL14,KlongvessaPRL19}.
Recent experimental and theoretical studies also showed that the self-spinning or
self-circling alone (of active spinners or rotors \cite{TaramaJPCM12,OhtaJPSJ17,
KrugerPRL16,NourhaniPRE16,KurzthalerSoftMatter17,ChepizhkoSoftMatter19}) can
generate spatiotemporal collective states among the interacting chiral particles
\cite{TsaiPRL05,YangPRE20,LiuPNAS20,NguyenPRL14,vanZuidenPNAS16,DasbiswasPNAS18,
LiebchenPRL17,LevisPRE19,FurthauerPRL13,KirchhoffEPL05,KaiserPRE13,LeiSciAdv19,
CapriniSoftMatter19,HanSciAdv20,SnezhkoCOCIS16,ShenSoftMatter19,YeoPRL15,
WangPRFluids19,FarhadiSoftMatter18,UchidaPRL10,LiaoSoftMatter18,YangPRE14}.
A typical example is the edge current flow of rotors generated at rigid boundary
walls; this edge mode induces the collective, unidirectional flow of neighboring
rotors which either decays into the confined sample interior of a liquid or gas phase
\cite{TsaiPRL05,YangPRE20,CapriniSoftMatter19,vanZuidenPNAS16,DasbiswasPNAS18,LiuPNAS20}
or causes the rotation of the circular sample in a crystalline or jammed phase
\cite{vanZuidenPNAS16,LiuPNAS20}.

Although much effort has been devoted to investigating either of these two
mechanisms of self-propulsion and self-spinning, effects of their mutual coupling
are much less explored. Also less understood is the corresponding crystallization
process, for which a statistical continuum
description that can access large length and time scales much beyond the restrictions
encountered in discrete particle-based approaches, has still been lacking.
Here, by introducing a continuum density-field description that
is nonpotential and nonvariational, we show that the coupling and competition between
self-propulsion and self-spinning result in a surprisingly rich behavior of
nonrelaxational dynamical crystallized states. They feature both translational
and rotational collective motion, governed by persistent
dynamics. Two types of transition for active crystalline patterns are identified,
i.e., bulk traveling-localization and interfacial localized-delocalized transitions,
each mediated by a crossover regime showing dynamical frustration of active
chiral particles. Of particular interest is the effect of the controlled
crystal-melt interfaces, leading to an emergent state of self-rotating crystallites
embedded in a homogeneous active melt, or self-shearing or wriggling flow of
crystalline layers.

The active system here is described by a local particle density variation field
$\psi$ and a local polar particle orientation field $\mathbf{P}$, the dynamics
of which are governed by
\begin{eqnarray}
  && \partial_t \psi =
  \nabla^2 \left [ \epsilon \psi + ( \nabla^2 + q_0^2 )^2 \psi - g \psi^2
    + \psi^3 \right ] - v_0 {\bm \nabla} \cdot \mathbf{P},~\quad \label{eq:psi}\\
  && \partial_t \mathbf{P} =
  \left ( \nabla^2 - D_r \right ) \left ( C_1 \mathbf{P} + C_4 |\mathbf{P}|^2 \mathbf{P}
  \right) - v_0 {\bm \nabla} \psi + \mathbf{M} \times \mathbf{P},\nonumber\\ \label{eq:P}
\end{eqnarray}
where $\epsilon$, together with the average density $\psi_0$, controls the transition
between homogeneous (liquid) and crystalline phases, $v_0$ is the self-propulsion
strength, $D_r$ is the rotational diffusion constant, and $\mathbf{M} = M \hat{z}$
represents the strength of self-spinning caused by an active torque.
When $\mathbf{M}=0$ we recover the previous active phase field crystal (PFC) description
\cite{MenzelPRL13,MenzelPRE14}.

Here we focus on $C_1 > 0$, counteracting any spontaneous polarization
by e.g., rotational diffusion, to preclude any explicit alignment interactions
and be consistent with related experiments. Our description can be derived
from a microscopic particle-based formulation and dynamical density functional theory
(DDFT) \cite{SM}. %as detailed in the Supplemental Material.
All the model parameters are rescaled and dimensionless, giving a diffusion time scale
and a spatial scale set by the periodicity of the ordered phase with $q_0=1$.

\begin{figure}
  \centerline{\includegraphics[width=0.5\textwidth]{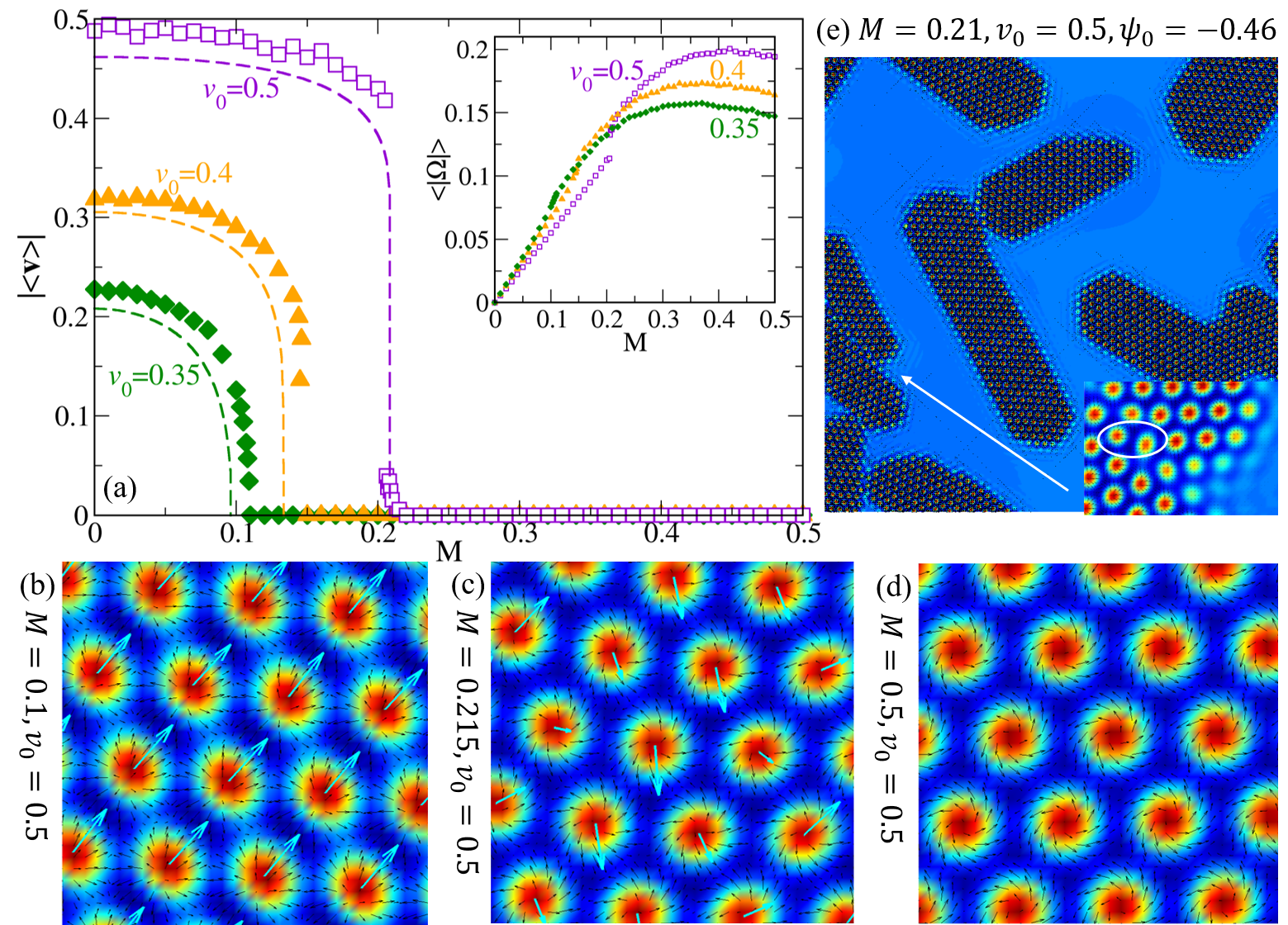}}
  \caption{(a) Magnitude of average density-peak velocity $|\langle\mathbf{v}\rangle|$
    as a function of $M$, for $\psi_0=-0.3$ and $v_0=0.35$, $0.4$, and $0.5$. Dashed curves
    represent analytic results of Eq.~(\ref{eq:vm}) assuming $\hat{q}_0 \simeq q_0=1$.
    Inset: $\langle|\Omega|\rangle$ measuring particle self-spinning.
    (b)--(d) Snapshots of parts of the $\psi$ profiles in three characteristic regimes,
    with the density peaks shown in red, their velocities $\mathbf{v}$ indicated
    by large arrows, and the polarization field $\mathbf{P}$ by fine arrows. 
    %Only the central portion of each simulation box is shown.
    (e) Snapshot of migrating crystallites in an active melt (see Movie S2).
    Inset: Enlarged portion of a two-grain impinged corner, with a penta-hepta dislocation.}
  \label{fig:v_M}
\end{figure}

It is noteworthy that Eqs.~(\ref{eq:psi}) and (\ref{eq:P}) could be reduced to a gradient,
relaxational form only when $v_0=M=0$ for passive particles, giving $\partial_t \psi =
\nabla^2 {\delta \mathcal{F}}/{\delta \psi}$ and $\partial_t \mathbf{P} = (\nabla^2 - D_r)
{\delta \mathcal{F}}/{\delta \mathbf{P}}$ with $\mathcal{F} =
\int d\mathbf{r} \{ \psi [ \epsilon + ( \nabla^2 + q_0^2 )^2 ] \psi/2 - g \psi^3/3
+ \psi^4/4 + C_1 |\mathbf{P}|^2/2 + C_4 |\mathbf{P}|^4/4 \}$, a combination of the
PFC free energy \cite{ElderPRL02,ElderPRB07,MkhontaPRL13,MkhontaPRL16} and a Landau
expansion of the polarization field. Thus for any active system with nonzero $v_0$,
the corresponding system dynamics is nonrelaxational, i.e., does not follow the
minimization of $\mathcal{F}$. We examine this nonpotential system through
a series of numerical simulations. Each starts either from a homogeneous state with
random initial conditions or from initially small crystalline nuclei, with system
sizes ranging from $128 \times 128$ to $2048 \times 2048$ grid points
(around $230$ to $60,000$ density peaks) subjected to periodic boundary
conditions. The system parameters are chosen as $(\epsilon, g, D_r, C_1, C_4) =
(-0.98, 0, 0.5, 0.2, 0)$, while values of $M$, $v_0$, and $\psi_0$ are varied to
control the competition between self-propulsion and self-spinning.

\begin{figure}
  \centerline{\includegraphics[width=0.5\textwidth]{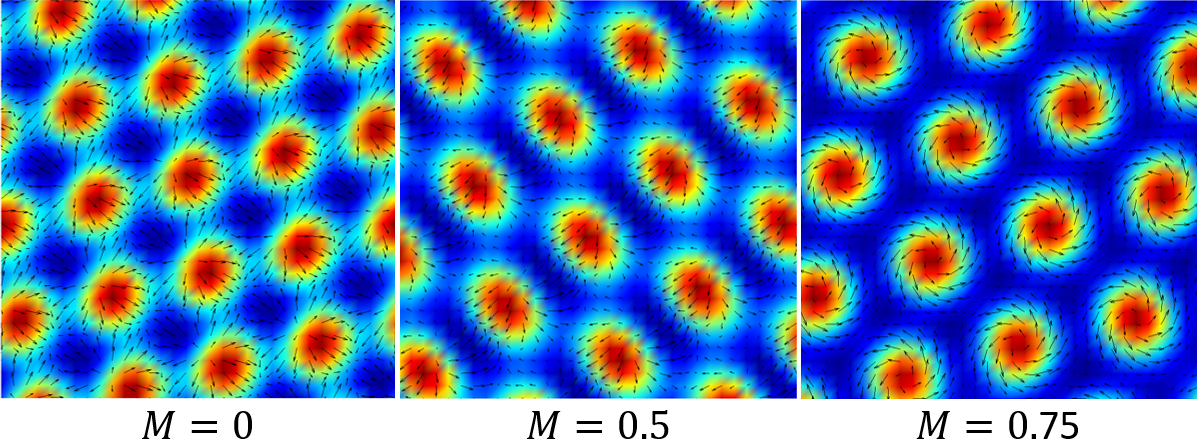}}
  \caption{Self-spinning-induced phase transformation of crystalline patterns with
  increasing $M$ (at $v_0=1$, $\psi_0=-0.4$).}
  \label{fig:phase_trans}
\end{figure}

Our simulations indicate that in the bulk state of active crystals three
characteristic regimes of system dynamics can be identified. As shown in
Fig.~\ref{fig:v_M}(a), the propelling-spinning competition leads to a sharp
transition between a unidirectionally traveling ordered
state driven by particle self-propulsion at large enough $v_0$ and small $M$
[Fig.~\ref{fig:v_M}(b)] and a localized or arrested crystalline state with vanishing
velocity $\mathbf{v}$ of each density peak at large enough $M$ [Fig.~\ref{fig:v_M}(d)].
The effect of active torque causes the self-spinning or localized self-circling of
individual chiral particles and hence the localization of each density peak,
which is consistent with previous DDFT results for noncrystallized states
\cite{HoellNJP17}. It thus induces the arrest of the whole pattern as observed here.
%The structural arrest also occurs for dislocation defects and vacancies appearing
%in large systems, which evolve and move very slowly or get pinned as a result of
%the localizing effect.
This traveling-localization transition can be accompanied by phase transformations
between ordered structures as induced by particle self-spinning when $M$ increases.
Examples include transformations from a traveling rhombic or distorted-hexagonal
structure to a localized hexagonal phase [Figs.~\ref{fig:v_M}(b)--\ref{fig:v_M}(d)],
or from a traveling-square to traveling-rhombic to localized-hexagonal
structures at large active drive $v_0$ (Fig.~\ref{fig:phase_trans}).

In a narrow crossover regime near the transition threshold ($M \sim M_c$)
the incompatibility between two dynamical effects dominated by self-propelled
translation and spinning-induced localization becomes explicit. When these two
competing dynamics are of similar degree, none of the corresponding optimal
collective behaviors can be achieved, leading to the local dynamic frustration
of active particles [see Fig.~\ref{fig:v_M}(c)] or a wavy, swinging translative
motion of the whole crystalline pattern characterized by alternative regions
of density peaks traveling at varying directions (see Supplemental Movie S1).
The balancing of translation and localization also leads to
a phenomenon of migrating crystallites. As shown in Fig.~\ref{fig:v_M}(e) and
Movie S2, during the evolution and coarsening of faceted crystallites or grains,
the density peaks are localized within each grain while the whole crystallites
travel within the coexisting medium of homogeneous melt, impinging and coalescing
with each other.

The above results can be further understood by rewriting Eq.~(\ref{eq:P}) in terms
of a local divergence field $S={\bm \nabla} \cdot \mathbf{P}$ and
the self-spinning field $\Omega=({\bm \nabla} \times \mathbf{P})_z/2$; for $C_4=0$,
\begin{eqnarray}
  && \partial_t S = C_1 \left ( \nabla^2 - D_r \right ) S - 2 M \Omega
  - v_0 \nabla^2 \psi, \label{eq:S}\\
  && \partial_t \Omega = C_1 \left ( \nabla^2 - D_r \right ) \Omega
  + \frac{1}{2} M S. \label{eq:Omega}
\end{eqnarray}
Equation (\ref{eq:Omega}) indicates that $S$ serves as an effective source for
nonzero $\Omega$ in the steady state, generating particle self-spinning locally,
as confirmed numerically in the inset of Fig.~\ref{fig:v_M}(a).
Nonzero $S$ also drives the propagation of density patterns when entering
Eq.~(\ref{eq:psi}), while its own source is in turn provided by the variation of
the density field [see $v_0\nabla^2 \psi$ in Eq.~(\ref{eq:S})]. Simultaneously,
the term $M \Omega$ in Eq.~(\ref{eq:S}) causes the damping of $S$
and plays the role of an inhibitor that hinders the particle migration,
leading to the effect of localization observed in the simulations.

Given the linear form of Eqs. (\ref{eq:S}) and (\ref{eq:Omega}),
it is straightforward to express the Fourier components of $S$ and $\Omega$ in
terms of those of density $\psi$ in the nonequilibrium steady state with a
constant pattern migration speed $v_m$. For the hexagonal phase,
in one-mode approximation
\begin{eqnarray}
  & v_m^2 =& \frac{2}{3\hat{q}_0^2} \Biglb \{ \hat{q}_0^2 v_0^2
  - 2[C_1^2(\hat{q}_0^2+D_r)^2-M^2] \nonumber\\
  && + \sqrt{\hat{q}_0^4 v_0^4 - 16M^2C_1^2(\hat{q}_0^2+D_r)^2} \Bigrb \}, \label{eq:vm}
\end{eqnarray}
if $v_m^2>0$ and $|M| \leq {\hat{q}_0^2 v_0^2}/[4C_1(\hat{q}_0^2+D_r)]$;
otherwise $v_m=0$. %(A detailed derivation will be given elsewhere.)
This analytic result indicates that there exists a critical threshold $M_c$
(or $v_{0c}$); when $|M| > M_c$ (or $v_0 < v_{0c}$) the active crystal is localized
with $v_m=0$. $\hat{q}_0$ is the selected wave number of the ordered pattern,
difficult to be determined analytically for a nonpotential system
\cite{CrossRMP93,CrossPRL95,HuangPRE07}.
Our simulations indicate that $\hat{q}_0$ is in the vicinity of $q_0$,
an approximation used in evaluating Eq.~(\ref{eq:vm}) as presented in
Fig.~\ref{fig:v_M}(a) without any parameter fitting. We find a reasonably
good agreement between analytic and numerical results for $v_m$,
with the deviations attributed to the employed one-mode approximation.

\begin{figure}
  \centerline{\includegraphics[width=0.5\textwidth]{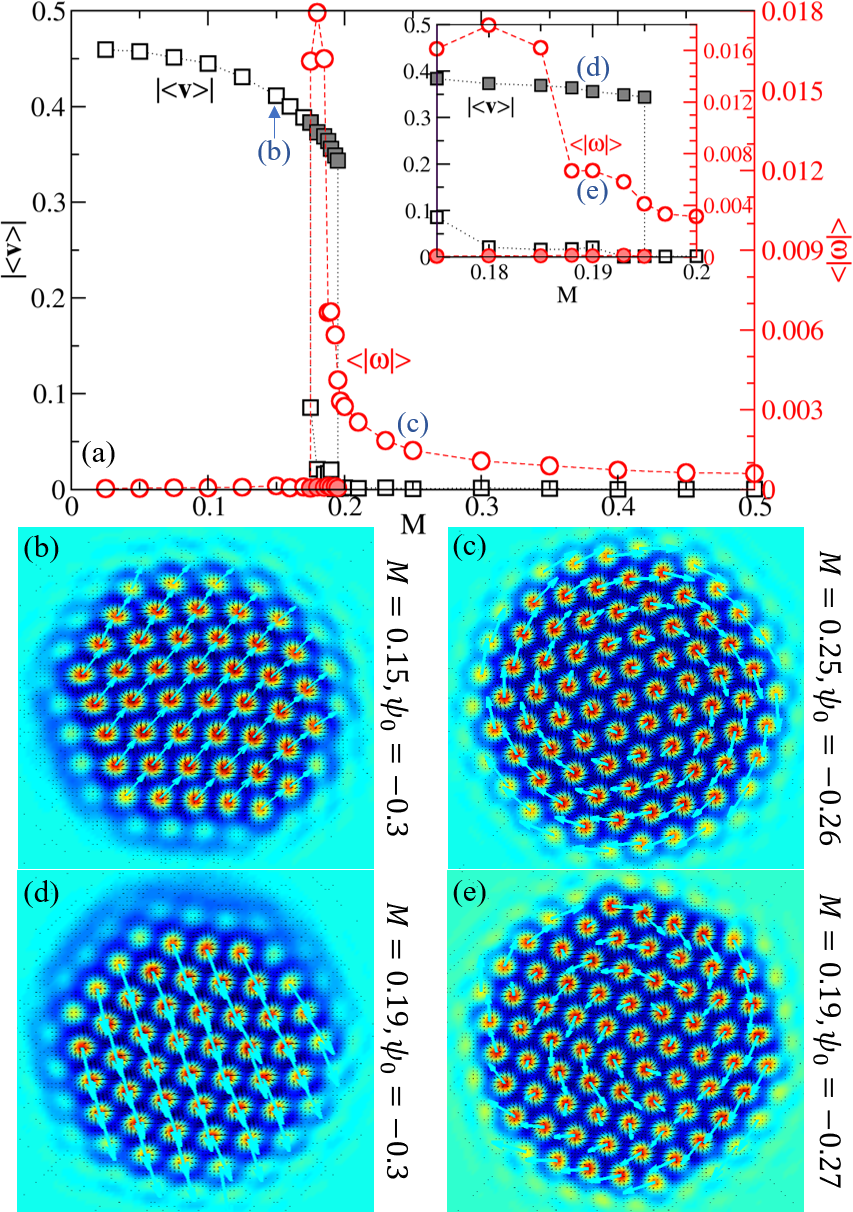}}
  \caption{(a) Averaged translational and rotational velocities
    ($|\langle\mathbf{v}\rangle|$ and $\langle|\omega|\rangle$ at $v_0=0.5$)
    as a function of $M$, containing translation- and rotation-dominated
    regimes for crystallites in a circular cavity and
    a transitional regime (shaded symbols) enlarged in the inset.
    (b)--(e) Snapshots of crystallite patterns simulated [as marked in (a)],
    with $\mathbf{v}$ vectors indicated.}
  \label{fig:cryst_rotating}
\end{figure}

Remarkably, the effect of chiral particle self-spinning, which leads to the
bulk structure localization identified above, can be utilized to
generate a further localized-delocalized transition in the interfacial state
and, consequently, a phenomenon of crystallite self-rotation or self-shearing.
The crystal-melt interface is set up through a sufficiently
steep gradient of the ``temperature''-type parameters $\epsilon$ and $C_1$
(corresponding to the spatial variation of, e.g., chemicals or heat sources
controlling homogeneous vs.\ crystalline phases in experiments). Our
obtained results are robust against specific realizations of the setup,
as verified in simulations. Below, we present results for a kink-type
two-phase profile
\begin{eqnarray}
  \epsilon &=& \frac{1}{2} \left \{ (\epsilon^{\rm in} + \epsilon^{\rm out}) +
  (\epsilon^{\rm out}-\epsilon^{\rm in}) \tanh \left [ (r-r_0)/\Delta \right ] \right \},\quad
\end{eqnarray}
with the same form for $C_1$. This represents a circular cavity of radius $r_0$,
enclosing a crystallite embedded in an outside coexisting homogeneous medium
of active melt. Similar kink-type profiles can be set up for other interfacial
geometries. Here we set $(\epsilon^{\rm in}, \epsilon^{\rm out}, C_1^{\rm in},
C_1^{\rm out}) = (-0.98, 0, 0.2, 1)$ and $\Delta$ around 1 to 5 grid spacings.

Simulation results for a circular cavity are presented in Fig.~\ref{fig:cryst_rotating},
showing three regimes of crystallite self-motion:
(i) self-translation dominated, (ii) self-rotation dominated, and (iii) the
transition between them, as a result of the competition between self-propelled
translative particle motion, interface-induced tangential motion of density peaks,
and localization through particle self-spinning. In regime (i) for small $M$,
high-density blocks constantly crystallize from the active melt at one side of the
cavity, propagate across it, and remelt into the homogeneous medium at the other side,
as seen in Fig.~\ref{fig:cryst_rotating}(b) and Movie S3. At the same time the whole
crystallite still rotates slowly, with small but nonzero averaged angular velocity
$\langle|\omega|\rangle$ [estimated as the orientation change rate
of $\langle \mathbf{v} \rangle$, given the dominance of translative motion,
with the center of rotation located outside the crystallite; see
Fig.~\ref{fig:cryst_rotating}(a)]. The maximum magnitude of the average translational
velocity $|\langle\mathbf{v}\rangle|$ (among realizations of different $\psi_0$)
decreases with increasing $M$, due to stronger localization through particle
self-spinning. At large enough $M$ [regime (ii)], persistent self-rotation of
the faceted crystallite about the cavity center is observed, as illustrated in
Fig.~\ref{fig:cryst_rotating}(c) and Movie S4. The direction of crystallite
rotation (clockwise) is opposite to that of individual particle self-spinning
(counterclockwise). The maximum averaged rotation rate $\langle|\omega|\rangle$
(here $|\omega| = |\mathbf{v} - \langle \mathbf{v} \rangle| /r$) is reduced at
larger $M$ with enhanced bulk localization.

This phenomenon of self-rotating crystallites can be understood by examining the
spatial variation of the polarization field $\mathbf{P}$.
At the cavity boundary the average polarization of a boundary density peak at the
melt side is weaker than at the inner crystalline side, generating a local
spatial gradient of $\mathbf{P}$ (and $\Omega$) that is suppressed asymmetrically
by its vicinity to the interface. Thus, the active drive at the inner side
of each boundary peak dominates over that of its outer side,
causing the corresponding self-flow of the interfacial crystalline layer.
It subsequently overcomes the localization of (i.e., delocalizes) interior
particles via collective dynamics and drives the overall self-rotation of the
crystallite. The direction of self-rotation follows the orientation of the net
polarization at the inner side of interfacial peaks and thus, interestingly, is
opposite to that determined by the chirality of individual self-spinning.
When the sign of $M$ is reversed, both directions of self-spinning
and self-rotation are reversed. %as confirmed in our simulations.
The mechanism here is different from that underlying the rotation or edge flows
of active spinners found in previous studies of no-flux rigid
boundary walls; there the chirality of the boundary/edge flow is the same as that
of the individual spinners or rotors \cite{vanZuidenPNAS16,DasbiswasPNAS18,
TsaiPRL05,YangPRE20,CapriniSoftMatter19,LiuPNAS20}.

In the transitional regime the crystallite shows
double-degenerate behaviors, one dominated by self-translation
and the other by self-rotation, as depicted in Figs.~\ref{fig:cryst_rotating}(d)
and \ref{fig:cryst_rotating}(e), respectively. In the latter, although the
overall crystallite self-rotates as driven by the interfacial layer,
some inner layers exhibit local frustration and even intermittent inverse
rotation [leading to regions of local self-shearing;
see Fig.~\ref{fig:cryst_rotating}(e) and Movie S5]. This reflects the
competition among self-propulsion, self-spinning-induced localization, and
interface-induced delocalization.

\begin{figure}
  \centerline{\includegraphics[width=0.5\textwidth]{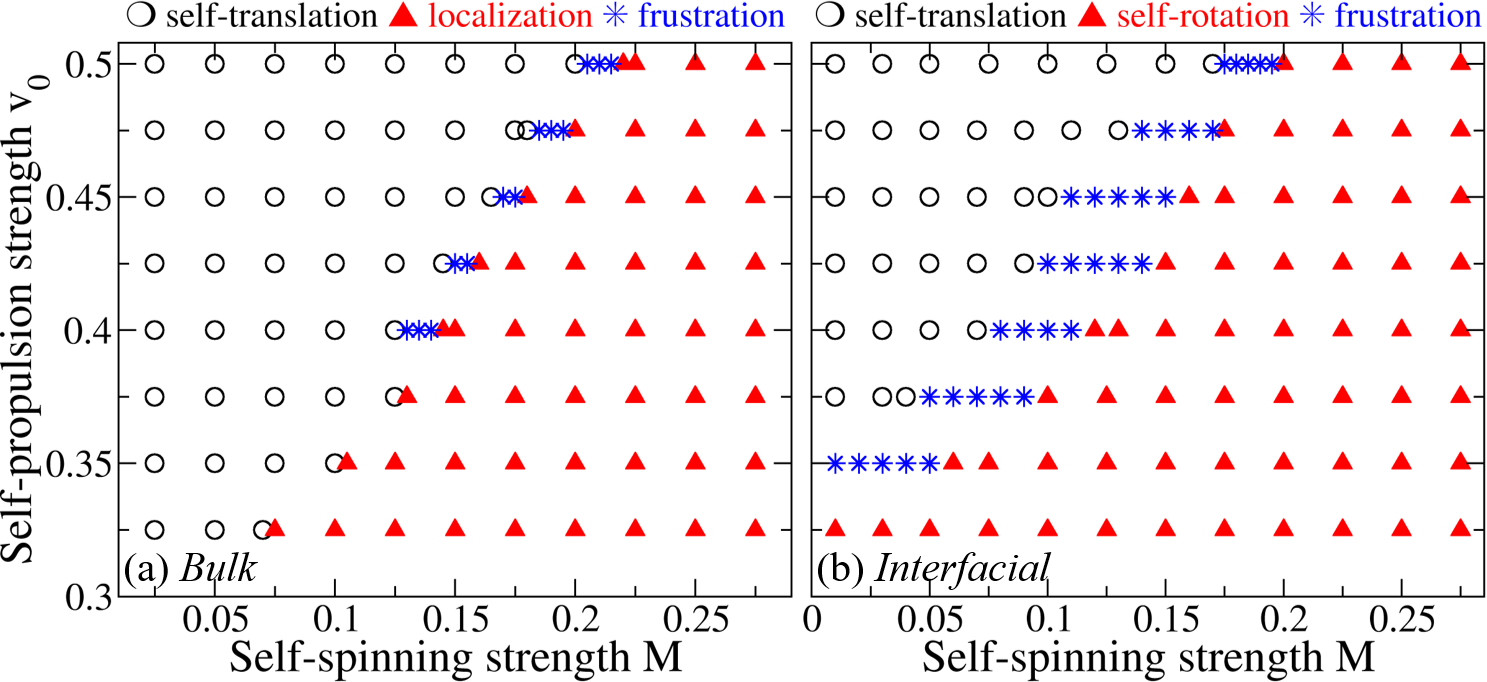}}
  \caption{Dynamical state diagrams in the $M$-$v_0$ space
  for three dynamical regimes in (a) bulk and (b) interfacial systems.}
  \label{fig:state_diagrams}
\end{figure}
  
The dynamical regimes identified for both bulk and circularly
interfacial systems are summarized in Fig.~\ref{fig:state_diagrams}, in terms of
$M$-$v_0$ state diagrams obtained from simulations across different values of $\psi_0$
that lead to hexagonal crystallized states. Major parts of the $M$-$v_0$ space are
occupied by the self-translation-dominated state %[at large (small) enough $v_0$ ($M$)] 
and the states of self-spinning-induced bulk localization [Fig.~\ref{fig:state_diagrams}(a)]
or interface-related crystallite self-rotation [Fig.~\ref{fig:state_diagrams}(b)].
The transitional regime, characterized by dynamical frustration in both cases,
is broader in Fig.~\ref{fig:state_diagrams}(b) that involves crystal-melt interfaces.

\begin{figure}
  \centerline{\includegraphics[width=0.5\textwidth]{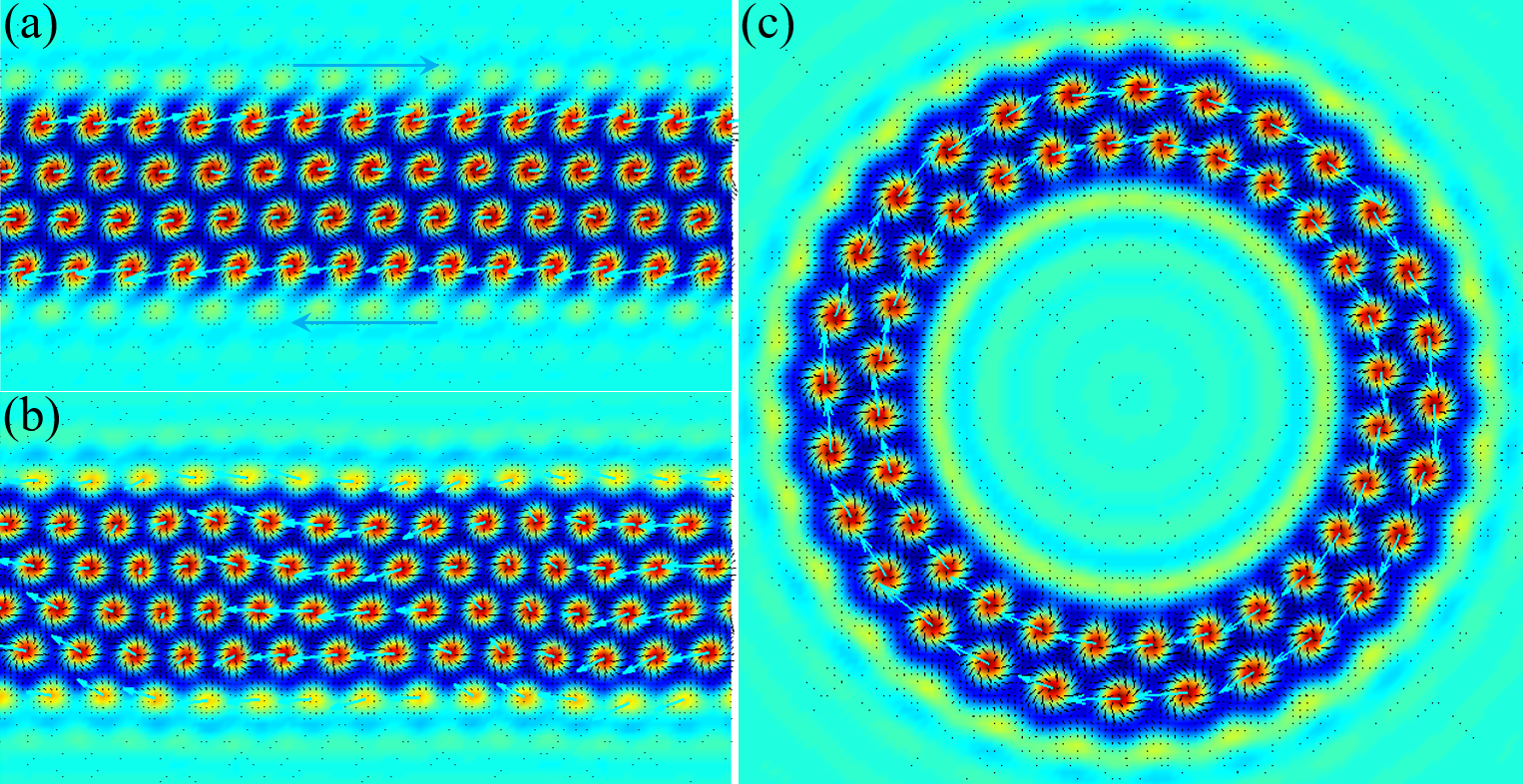}}
  \caption{Interface-induced self-motion of crystalline layers
    at $v_0=0.5$, including (a) self-shearing ($M=0.5$, $\psi_0=-0.2$) and
    (b) frustrated motion ($M=0.19$, $\psi_0=-0.19$) for a slab crystallite, and
    (c) self-rotation of a ring crystallite ($M=0.25$, $\psi_0=-0.26$).}
  \label{fig:self_shearing}
\end{figure}

The interface-induced driving mechanism of crystalline layers should apply
to any geometry of nonrigid (soft) crystal-liquid boundaries or edges.
In the example of a slab crystallite [Fig.~\ref{fig:self_shearing}(a)],
the top and bottom interfacial layers are expected to be driven towards opposite
directions due to their inverse crystal-to-liquid interface orientations,
leading to self-shearing of the crystallite as verified in our simulations
[see Fig.~\ref{fig:self_shearing}(a) and Movie S6]. Conversely,
in the transitional regime local frustration of density peaks
occurs as a consequence of the comparable strengths of self-propulsion and
self-spinning, as seen in Fig.~\ref{fig:self_shearing}(b). This leads to either
fluctuating modes of traveling crystalline layers (Movie S7) or even a
snaking/worming type of layer flow (Movie S8).

Finally, similar consequences of opposing interfacial orientations can be manifested
in a ring-like configuration, as shown in Fig.~\ref{fig:self_shearing}(c)
and Movie S9. In many cases, although the whole ring-shaped crystallite still
rotates collectively as driven by the outermost circular layer with longest
perimeter [Fig.~\ref{fig:self_shearing}(c)], its angular velocity is significantly
reduced due to the hindrance by the counteracting drive of the innermost annulus.
The scenario of self-shearing occurs when the two interfacial drives
are of comparable strength, for which the outer and inner crystalline layers
rotate to opposite directions (Movie S9).

In summary, we have analyzed and predicted the collective behavior of
spatially ordered structures featuring both active propulsion and active
rotation. The interplay between individual particle self-propulsion and
self-spinning during crystallization results in
various novel states of collective and persistent dynamics that are enabled
by the nonrelaxational nature of the active system. The competition leads to
a traveling-frustration-localization transition in active crystals with
increasing strength of self-spinning, which also induces a transformation
between ordered phases as a result of pattern selection. A breaking of the
localization and structural arrest occurs for interfacial states, revealing
persistently dynamical states of self-rotating crystallites and self-shearing
or self-flowing crystalline layers. The direction of crystallite self-rotation
or layer propagation is opposite to that given by the chirality of the individual
self-spinning particles, an effect caused by the crystal-melt interface-induced
spatial variation of local polarization and the subsequent edge-originated
delocalization and collective motion of active particles.

These predictions open new possibilities to explore the emergence of novel
dynamical phenomena and unveil complex mechanisms underlying a wide variety
of nonequilibrium active systems governed by persistent, nonrelaxational dynamics.
Although here we focused on a dry environment, the results and
mechanisms identified above will in many cases still apply to leading order in
additional fluid surroundings \cite{SM}. It should even be possible to
disentangle the effects of both environments in an experiment when surrounding
granular or colloidal spinners by a viscous fluid with varying viscosity for
the tuning of hydrodynamic couplings.
Our results can be realized and verified in various experimental setups, such as
a collective of granular rotors \cite{YangPRE20,LiuPNAS20,ScholzNatCommun18},
light-controlled anisotropic colloidal Janus particles \cite{KummelPRL13} and
colloidal molecules \cite{SchmidtJCP19}, or self-propelled particles equipped
with magnetic dipole moments \cite{BarabanACSNano12,TaukulisJMMM14} to perform
active spinning under a magnetic field.

\begin{acknowledgments}
  Z.-F.H. acknowledges support from the National Science Foundation under Grant
  No.~DMR-1609625. A.M.M. thanks the German Research Foundation (DFG) for support through
  the Heisenberg Grant ME 3571/4-1. H.L. was supported by the German Research Foundation
  (DFG) within project LO 418/20-2.
\end{acknowledgments}

\bibliography{activePFC2D_rotating_references}

\end{document}

%%% Title 
\title{Dynamical Crystallites of Active Chiral Particles \\
---\\
Supplemental Material}

%%% Authors
\author{Zhi-Feng Huang}
\affiliation{Department of Physics and Astronomy, Wayne State University, Detroit, Michigan 48201, USA}
\author{Andreas M. Menzel}
\affiliation{Institut f\"ur Theoretische Physik II, Heinrich-Heine-Universit\"at D\"usseldorf,
  D-40225 D\"usseldorf, Germany}
\author{Hartmut L\"owen}
\affiliation{Institut f\"ur Theoretische Physik II, Heinrich-Heine-Universit\"at D\"usseldorf,
  D-40225 D\"usseldorf, Germany}
%

%%% Date
\date{\today}

\begin{abstract}
  In this supplemental material, we provide a more microscopic basis of the active phase field crystal
  (PFC) description introduced in Eqs.~(1) and (2) of the main text. Frequently, these equations are
  directly used as an input to corresponding evaluations. However, a derivation and motivation from
  a microscopic picture is possible, as we summarize below. We start from a particle-based picture of
  active microscopic objects that besides pure translational self-propulsion feature self-spinning
  as well. It is demonstrated how Eqs.~(1) and (2) of the main text follow from the corresponding
  statistical continuum representation of the discrete microscopic picture under suitable rescaling.
  In addition, effects of hydrodynamic interactions on the active PFC description are briefly discussed.
\end{abstract}

\maketitle

\subsection{Details on microscopic derivation of the active PFC description}

We start from a system of $N$ microscopic spherical active particles located at positions
$\mathbf{r}_1(t),...,\mathbf{r}_N(t)$ and featuring marked orientations given by unit vectors
$\mathbf{\hat{u}}_1(t),...,\mathbf{\hat{u}}_N(t)$, where $t$ denotes time. Considering overdamped
dynamics of the Langevin type, the time evolution of the positions and orientations of the particles
is determined by their deterministic velocities $\mathbf{v}_{\mathrm{det},i}$ and angular velocities
$\bm{\omega}_{\mathrm{det},i}$ that can be calculated from the deterministic forces and torques acting
on them, plus the influence of stochastic translational noise $\bm{\xi}_{\mathrm{tr},i}$ and stochastic
orientational noise $\bm{\xi}_{\mathrm{rot},i}$ acting on each particle ($i=1,...,N$). The stochastic
noises are typically viewed as thermal in origin, delta-correlated in time, and Gaussian in
distribution. Originating from thermal equilibrium fluctuations, the fluctuation-dissipation relation
sets its strength \cite{kubo1991statistical}. During an infinitesimal time step $\mathrm{d}t$, the
positional and orientational changes are governed by 
\begin{eqnarray}
\label{eq_dr}
\mathrm{d}\mathbf{r}_i &=& \mathbf{v}_{\mathrm{det},i}\,\mathrm{d}t + \bm{\xi}_{\mathrm{tr},i}\sqrt{\mathrm{d}t}, \\
\label{eq_du}
\mathrm{d}\mathbf{\hat{u}}_i &=& \bm{\omega}_{\mathrm{det},i}\,\mathrm{d}t\times\mathbf{\hat{u}}_i
+ \bm{\xi}_{\mathrm{rot},i}\sqrt{\mathrm{d}t}\times\mathbf{\hat{u}}_i. 
\end{eqnarray}

Because of the dynamics of the individual particles, the probability density $\mathcal{P}(\mathbf{r}_1,...,
\mathbf{r}_N,\mathbf{\hat{u}}_1,...,\mathbf{\hat{u}}_N,t)$ for finding the system at a time $t$ in a
certain configuration $\mathbf{r}_1,...,\mathbf{r}_N,\mathbf{\hat{u}}_1,...,\mathbf{\hat{u}}_N$ evolves
with time. Through textbook procedures \cite{risken1996the,zwanzig2001nonequilibrium}, by averaging
over the noise, Eqs.~(\ref{eq_dr}) and (\ref{eq_du}) can be transformed into a statistical conservation
equation for the time evolution of $\mathcal{P}$. Since we consider overdamped dynamics, the time
evolution of $\mathcal{P}$ is solely driven by the velocities $\mathbf{v}_i$ and angular velocities
$\bm{\omega}_i$ of the particles (with $i=1,...,N$). Along these lines, the dynamics of $\mathcal{P}$
thus follows the conservation equation \cite{doi1986theory}
\begin{equation}
\label{eq_P}
\frac{\partial \mathcal{P}}{\partial t} = {}-\sum_{i=1}^N \Big[ \nabla_{\mathbf{r}_i}\cdot(\mathbf{v}_i\mathcal{P})
  +(\mathbf{\hat{u}}_i\times\nabla_{\mathbf{\hat{u}}_i})\cdot(\bm{\omega}_i\mathcal{P})\Big]. 
\end{equation}
 
Both the stochastic noise acting on the particles as well as the deterministic forces and torques control
the particle velocities $\mathbf{v}_i$ and angular velocities $\bm{\omega}_i$ ($i=1,...,N$). Here we
consider deterministic potential interactions between the particles, deriving from the overall potential
$U(\mathbf{r}_1,...,\mathbf{r}_N,\mathbf{\hat{u}}_1,...,\mathbf{\hat{u}}_N)$. Since we are working towards
the probabilistic conservation equation for $\mathcal{P}$ in Eq.~(\ref{eq_P}), instead of explicitly working
out the influence of stochastic noises in Eqs.~(\ref{eq_dr}) and (\ref{eq_du}), an equally valid approach
at this stage is to supplement $U$ by an entropic potential $k_{\mathrm{B}}T\ln \mathcal{P}$ \cite{doi1986theory},
with $k_{\mathrm{B}}$ denoting the Boltzmann constant and $T$ the absolute temperature. This entropic potential
implies a diffusional behavior as obtained by explicitly averaging the noise terms in Eqs.~(\ref{eq_dr})
and (\ref{eq_du}). Considering active particles self-propelling and self-spinning on a substrate, the
resulting forces and torques acting on the particles are given by
\begin{eqnarray}
\label{eq_F}
\mathbf{F}_i &=& {}-\nabla_{\mathbf{r}_i}\left(U+k_{\mathrm{B}}T\ln \mathcal{P}\right)
+ F^{\mathrm{a}}\mathbf{\hat{u}}_i, \\
\label{eq_T}
\mathbf{T}_i &=& {}-\mathbf{\hat{u}}_i\times\nabla_{\mathbf{\hat{u}}_i}\left(U+k_{\mathrm{B}}T\ln \mathcal{P}\right)
+ \mathbf{T}^{\mathrm{a}}_i, 
\end{eqnarray}
with $i=1,...,N$. Here $F^{\mathrm{a}}$ is the strength of the active self-propulsion force driving particle
$i$ along its orientation $\mathbf{\hat{u}}_i$, and $\mathbf{T}^{\mathrm{a}}_i$ is the active torque driving
the self-spinning of each particle $i$. Both $F^{\mathrm{a}}$ and $\mathbf{T}^{\mathrm{a}}$ are assumed to be
constant and identical for all the particles in the present investigation. 

Denoting by $\mu^{\mathrm{t}}$ and $\mu^{\mathrm{r}}$ the translational and rotational mobilities, respectively,
the velocity and angular velocity of particle $i$ are given by
\begin{eqnarray}
\label{eq_mut}
\mathbf{v}_i&=&\mu^{\mathrm{t}}\mathbf{F}_i,\\
\label{eq_mur}
\bm{\omega}_i&=&\mu^{\mathrm{r}}\mathbf{T}_i.
\end{eqnarray}
%$\mathbf{v}_i=\mu^{\mathrm{t}}\mathbf{F}_i$ and $\bm{\omega}_i=\mu^{\mathrm{r}}\mathbf{T}_i$
When substituting %these expression together with Eqs.~(\ref{eq_F}) and (\ref{eq_T}) 
Eqs.~(\ref{eq_F})--(\ref{eq_mur}) into Eq.~(\ref{eq_P}), we find
\begin{widetext}
\begin{equation}
\label{eq_Pdyn}
\frac{\partial \mathcal{P}}{\partial t} =
\sum_{i=1}^N\bigg\{ \nabla_{\mathbf{r}_i}\cdot
\Big( \mu^{\mathrm{t}}\left[\nabla_{\mathbf{r}_i}\left(U+k_{\mathrm{B}}T\ln \mathcal{P}\right)
  - F^{\mathrm{a}}\mathbf{\hat{u}}_i\right] \Big)\mathcal{P}
+ (\mathbf{\hat{u}}_i\times\nabla_{\mathbf{\hat{u}}_i})\cdot
\Big( \mu^{\mathrm{r}}\left[\mathbf{\hat{u}}_i\times\nabla_{\mathbf{\hat{u}}_i}
  \left(U+k_{\mathrm{B}}T\ln \mathcal{P}\right) - \mathbf{T}^{\mathrm{a}}_i\right] \Big)\mathcal{P}
\bigg\}.
\end{equation}
Following the standard procedure and assuming identical particles, from the above equation we can then
identify the dynamic equations for the $n$-particle densities $\rho^{(n)}$ ($1\leq n\leq N$) through
\begin{equation}
\label{eq_nbody}
\rho^{(n)}(\mathbf{r}_1,...,\mathbf{r}_n,\mathbf{\hat{u}}_1,...,\mathbf{\hat{u}}_n,t)=
\frac{N!}{(N-n)!}\int\mathrm{d}\mathbf{r}_{n+1}...\mathrm{d}\mathbf{r}_N \:
\mathrm{d}\mathbf{\hat{u}}_{n+1}...\mathrm{d}\mathbf{\hat{u}}_N \,
\mathcal{P}(\mathbf{r}_1,...,\mathbf{r}_N,\mathbf{\hat{u}}_1,...,\mathbf{\hat{u}}_N,t).
\end{equation}
\end{widetext}

In the following we confine ourselves to the two-dimensional (2D) subspace that is defined by the planar
substrate on which the active particles are self-propelling and self-spinning. In principle, most of the
steps described below follow analogously in three dimensions, but the terms involving the rotational
operator $\mathbf{\hat{u}}_i\times\nabla_{\mathbf{\hat{u}}_i}$ can be substantially simplified in two
dimensions. Denoting by $\mathbf{\hat{z}}$ the normal to the 2D plane and within this plane using the
angle $\varphi_i$ to represent the orientation $\mathbf{\hat{u}}_i$ of the $i$th particle measured from a
fixed axis, we have $\mathbf{\hat{u}}_i\times\nabla_{\mathbf{\hat{u}}_i}=\mathbf{\hat{z}}\,\partial_{\varphi_i}$.
Moreover, for particles self-spinning within the plane we set $\mathbf{T}^{\mathrm{a}}=T^{\mathrm{a}}\mathbf{\hat{z}}$,
with $T^{\mathrm{a}}$ the constant magnitude of the active torque. 

Next, applying the operation defined by Eq.~(\ref{eq_nbody}) to Eq.~(\ref{eq_Pdyn}) and denoting
$\rho^{(1)}(\mathbf{r}_1,\mathbf{\hat{u}}_1,t)$ as $\rho(\mathbf{r},\mathbf{\hat{u}},t)$, we find
the dynamic equation for the one-particle density $\rho$ as
\begin{eqnarray}
\label{eq_rho}
\frac{\partial\rho}{\partial t}
&=&
\mu^{\mathrm{t}}\,\nabla\cdot\left[ \widetilde{\mathcal{P}\nabla V}+ \rho\nabla\Phi^{\mathrm{ext}}
  + k_{\mathrm{B}}T\nabla\rho-F^{\mathrm{a}}\rho\mathbf{\hat{u}} \right]
\nonumber\\
&&{}+ \mu^{\mathrm{r}}\,\partial_{\varphi}\left[ \widetilde{\mathcal{P}\partial_{\varphi}\! V}
  + \rho\,\partial_{\varphi}\Phi^{\mathrm{ext}}+ k_{\mathrm{B}}T\partial_{\varphi}\rho-{T}^{\mathrm{a}}\rho \right],
\nonumber\\
\end{eqnarray}
where we have abbreviated
\begin{eqnarray}
\widetilde{\mathcal{P}\nabla V}
&=&
N\int \mathrm{d}\mathbf{r}_{2}...\mathrm{d}\mathbf{r}_N \:
\mathrm{d}\mathbf{\hat{u}}_{2}...\mathrm{d}\mathbf{\hat{u}}_N \mathcal{P}\nabla V,
\\
\widetilde{\mathcal{P}\partial_{\varphi} V}
&=&
N\int \mathrm{d}\mathbf{r}_{2}...\mathrm{d}\mathbf{r}_N \:
\mathrm{d}\mathbf{\hat{u}}_{2}...\mathrm{d}\mathbf{\hat{u}}_N \mathcal{P}\partial_{\varphi} \!V,
\end{eqnarray}
and split $U$ into the potential $V$ for mutual particle interactions and the possible influence
of an external potential ${\Phi}^{\mathrm{ext}}$. If $V$ only includes pairwise interactions, $U$ reads
\begin{equation}
\label{eq_UVPhi}
U=\sum_{i=1}^{N}\sum_{j>i}^N V(\mathbf{r}_i,\mathbf{r}_j,\mathbf{\hat{u}}_i,\mathbf{\hat{u}}_j) 
+\sum_{i=1}^N \Phi^{\mathrm{ext}}(\mathbf{r}_i,\mathbf{\hat{u}}_i),
\;
\end{equation}
and
\begin{eqnarray}
\label{eq_PnablaV}
\widetilde{\mathcal{P}\nabla V}
&=&
\int \mathrm{d}\mathbf{r}'\mathrm{d}\mathbf{\hat{u}}'\,
\rho^{(2)}(\mathbf{r},\mathbf{r}',\mathbf{\hat{u}},\mathbf{\hat{u}}')
\nabla V(\mathbf{r},\mathbf{r}',\mathbf{\hat{u}},\mathbf{\hat{u}}'),
\qquad 
%\nonumber\\
%&&{}+\rho(\mathbf{r},\mathbf{\hat{u}})\nabla\Phi^{\mathrm{ext}}(\mathbf{r},\mathbf{\hat{u}}),
\\%[.1cm]
\label{eq_PvarphiV}
\widetilde{\mathcal{P}\partial_{\varphi} \!V}
&=&
\int \mathrm{d}\mathbf{r}'\mathrm{d}\mathbf{\hat{u}}'\,
\rho^{(2)}(\mathbf{r},\mathbf{r}',\mathbf{\hat{u}},\mathbf{\hat{u}}')
\partial_{\varphi} \!V(\mathbf{r},\mathbf{r}',\mathbf{\hat{u}},\mathbf{\hat{u}}').
%\nonumber\\
%&&{}+\rho(\mathbf{r},\mathbf{\hat{u}})\partial_{\varphi}\Phi^{\mathrm{ext}}(\mathbf{r},\mathbf{\hat{u}}).
\end{eqnarray}

As a central step, we need to find a closure for these expressions in terms of $\rho$.
One strategy is provided by dynamical density functional theory (DDFT)
\cite{marconi1999dynamic,marconi2000dynamic, archer2004dynamical}, which has been successfully
developed and evaluated to describe, amongst many others, the dynamics of self-propelled particles
\cite{wensink2008aggregation,menzel2016dynamical, hoell2017dynamical,hoell2018particle,hoell2019multi}.
It applies exact equilibrium relations for the density to the instant state of the system density
at each moment of time $t$ during the nonequilibrium dynamics. This adiabatic approximation is
supported by the overdamped dynamics of the system. Here we provide a short summary of the procedure. 

Classical density functional theory exploits that the grand potential $\Omega$ in equilibrium shows
an extremum as a functional of the density \cite{evans2010density,lowen2010density}, 
\begin{equation}
\frac{\delta\Omega}{\delta\rho}=0. 
\end{equation}
Covering the effect of a possible chemical potential by the external potential, this implies 
\begin{equation}
\label{eq_sumF}
\frac{\delta\mathcal{F}^{\mathrm{id}}}{\delta\rho} +\frac{\delta\mathcal{F}^{\mathrm{exc}}}{\delta\rho}
+\frac{\delta\mathcal{F}^{\mathrm{ext}}}{\delta\rho}=0, 
\end{equation}
where $\mathcal{F}^{\mathrm{id}}=k_{\mathrm{B}}T\int\mathrm{d}\mathbf{r}\,\mathrm{d}\mathbf{\hat{u}}\,
\rho\left[\ln(\lambda^2\rho)-1\right]$ is the free energy associated with noninteracting particles
(i.e., the contribution from entropy) and $\lambda$ denotes the thermal
de Broglie wavelength. $\mathcal{F}^{\mathrm{exc}}$ accounts for the additional contributions to the free
energy resulting from interparticle interactions, while $\mathcal{F}^{\mathrm{ext}}=\int\mathrm{d}\mathbf{r}\,
\mathrm{d}\mathbf{\hat{u}}\,\rho\,\Phi^{\mathrm{ext}}$ includes into the free energy the effect of
the external potential. Thus, we get from Eq.~(\ref{eq_sumF}) 
\begin{equation}
\label{eq_Phi}
\Phi^{\mathrm{ext}}=-k_{\mathrm{B}}T\ln(\lambda^2\rho)-\frac{\delta\mathcal{F}^{\mathrm{exc}}}{\delta\rho}. 
\end{equation}

Simultaneously, in equilibrium ${\partial\rho}/{\partial t}=0$, $F^{\mathrm{a}}=0$, and ${T}^{\mathrm{a}}={0}$
in Eq.~(\ref{eq_rho}). Thus, the remaining translational and rotational parts in the first and second lines
on the right-hand side of Eq.~(\ref{eq_rho}) need to vanish (separately). Together with Eq.~(\ref{eq_Phi}), we obtain
\begin{eqnarray}
\label{eq_PnablaV_Fext}
\widetilde{\mathcal{P}\nabla V} &=& \rho\,\nabla\frac{\delta\mathcal{F}^{\mathrm{exc}}}{\delta\rho}, 
\\
\label{eq_PvarphiV_Fext}
\widetilde{\mathcal{P}\partial_{\varphi}\! V} &=& \rho\,\partial_{\varphi}\frac{\delta\mathcal{F}^{\mathrm{exc}}}{\delta\rho}. 
\end{eqnarray}
We remark that a similar functional form can be found if we apply the mean-field approximation
$\rho^{(2)}(\mathbf{r},\mathbf{r}',\mathbf{\hat{u}},\mathbf{\hat{u}}')
=\rho(\mathbf{r},\mathbf{\hat{u}})\rho(\mathbf{r}',\mathbf{\hat{u}}')$ in Eqs.~(\ref{eq_PnablaV})
and (\ref{eq_PvarphiV}). Then, the role of the expression $\delta\mathcal{F}^{\mathrm{exc}}/\delta\rho$
in Eqs.~(\ref{eq_PnablaV_Fext}) and (\ref{eq_PvarphiV_Fext}) is taken by the mean-field potential
\begin{equation}
  \int \mathrm{d}\mathbf{r}'\mathrm{d}\mathbf{\hat{u}}'\, \rho(\mathbf{r}',\mathbf{\hat{u}}')
  V(\mathbf{r},\mathbf{r}',\mathbf{\hat{u}},\mathbf{\hat{u}}'). 
\end{equation}

Using Eqs.~(\ref{eq_PnablaV_Fext}) and (\ref{eq_PvarphiV_Fext}) in Eq.~(\ref{eq_rho}), we have
derived the conservation equation for the one-particle density in terms of DDFT. Without taking into
account the role of an external potential as in the main text, i.e., setting $\Phi^{\mathrm{ext}}=0$
from now on, which leads to $\mathcal{F}=\mathcal{F}^{\mathrm{id}}+\mathcal{F}^{\mathrm{exc}}$, this
fundamental DDFT equation reads 
\begin{eqnarray}
\label{eq_ddft}
\frac{\partial\rho}{\partial t}
&=&
\mu^{\mathrm{t}}\,\nabla\cdot\left ( \rho\,\nabla\frac{\delta\mathcal{F}}{\delta\rho} \right )
-\mu^{\mathrm{t}}F^{\mathrm{a}}\,\nabla\cdot\mathbf{\hat{u}}\rho 
\nonumber\\
&&{}+
\mu^{\mathrm{r}}\,\partial_{\varphi} \left ( \rho\,\partial_{\varphi}\frac{\delta\mathcal{F}}{\delta\rho}
\right ) -\mu^{\mathrm{r}}\,{T}^{\mathrm{a}}\,\partial_{\varphi}\rho .
\end{eqnarray}

We now perform an expansion of the density $\rho$ with respect to its orientational dependence
\cite{lowen2010phase, wittkowski2010derivation,wittkowski2011polar}, 
\begin{equation}
  \rho(\mathbf{r},\mathbf{\hat{u}},t)=\bar{\rho}+\bar{\rho}\,\tilde{\phi}_1(\mathbf{r},t)
  +\bar{\rho}\,\mathbf{\hat{u}}\cdot\tilde{\mathbf{P}}(\mathbf{r},t)+...,
\end{equation}
and use this expansion to express Eq.~(\ref{eq_ddft}) in terms of $\tilde{\phi}_1$ and
$\tilde{\mathbf{P}}$, where we apply the approximation of constant mobility.
$\bar{\rho}$ is the average density. Moreover, we introduce the translational diffusion
constant $\tilde{D}=k_{\mathrm{B}}T\mu^{\mathrm{t}}$, the rotational diffusion constant
$\tilde{D}_r=k_{\mathrm{B}}T\mu^{\mathrm{r}}$, and the effective velocity of self-propulsion
$v=\mu^{\mathrm{t}}F^{\mathrm{a}}$. Additionally, we determine via integrations
$\int\mathrm{d}\mathbf{\hat{u}}$ and $\int\mathrm{d}\mathbf{\hat{u}}\,\mathbf{\hat{u}}$
the zeroth- and first-order orientational moments of Eq.~(\ref{eq_ddft}) to find the dynamic
equations for $\tilde{\phi}_1$ and $\tilde{\mathbf{P}}$. Defining $\beta=1/k_{\mathrm{B}}T$ as usual,
this guides us to
\begin{eqnarray}
\label{eq_tildephi1}
\frac{\partial(\bar{\rho}\tilde{\phi}_1)}{\partial t} &=&
\frac{\beta\tilde{D}\bar{\rho}}{2\pi}\nabla^2\frac{\delta\mathcal{F}}{\delta(\bar{\rho}\tilde{\phi}_1)}
-\frac{v}{2}\nabla\cdot(\bar{\rho}\tilde{\mathbf{P}}),
\\
\frac{\partial(\bar{\rho}\tilde{\mathbf{P}})}{\partial t} &=&
\frac{\beta\tilde{D}\bar{\rho}}{\pi}\nabla^2\frac{\delta\mathcal{F}}{\delta(\bar{\rho}\tilde{\mathbf{P}})}
-\frac{\beta\tilde{D}_r\bar{\rho}}{\pi}\frac{\delta\mathcal{F}}{\delta(\bar{\rho}\tilde{\mathbf{P}})}
-{v}\nabla(\bar{\rho}\tilde{\phi}_1)
\nonumber\\
&&
\label{eq_tildeP}
{}+\tilde{\mathbf{M}}\times(\bar{\rho}\tilde{\mathbf{P}}).
\end{eqnarray} 
In the last term, for compact notation we implicitly involve the direction $\mathbf{\hat{z}}$ normal
to the plane confining the particles. More precisely, here we have defined the effective angular
velocity of self-spinning $\tilde{\mathbf{M}}=\mu^{\mathrm{r}}T^{\mathrm{a}}\pi\,\mathbf{\hat{z}}$.

At this stage, our theory would be complete. However, exact expressions for $\mathcal{F}^{\mathrm{exc}}$
that contributes to $\mathcal{F}=\mathcal{F}^{\mathrm{id}}+\mathcal{F}^{\mathrm{exc}}$ in
Eqs.~(\ref{eq_tildephi1}) and (\ref{eq_tildeP}) are hardly known, except for some special model
cases. Therefore, approximations for $\mathcal{F}$ are necessary. We split the free energy
functional into a purely ${\phi}$-dependent part $\tilde{\mathcal{F}}_{\mathrm{pfc}}$ (where
$\phi=\phi_0+\bar{\rho}\tilde{\phi}_1$ relates to the spatial density variations) and
a purely $\tilde{\mathbf{P}}$-dependent part $\tilde{\mathcal{F}}_{\tilde{\mathbf{P}}}$,
i.e., $\mathcal{F}=\tilde{\mathcal{F}}_{\mathrm{pfc}}+\tilde{\mathcal{F}}_{\tilde{\mathbf{P}}}$. 
First, for the $\phi$-dependent part we choose the well-established phase field crystal (PFC)
description \cite{ElderPRL02,elder2004modeling,elder2007phase}
\begin{equation}
  \label{eq_Fpfc}
\tilde{\mathcal{F}}_{\mathrm{pfc}}=
\int\!\mathrm{d}\mathbf{r}\,\bigg \{
\frac{1}{2}\phi\left[a\,\Delta T+\lambda(\tilde{q}_0^2+\nabla^2)^2\right]\phi
-\frac{\tilde{g}}{3}\phi^3+\frac{u}{4}\phi^4\bigg \},
\end{equation}
%supplemented by the cubic term.
%\begin{equation}
%\label{eq_Fpfc}
%\tilde{\mathcal{F}}_{\mathrm{pfc}}=
%\int\!\mathrm{d}\mathbf{r}\,\bigg\{
%\frac{1}{2}\phi\left[a\,\Delta T+\lambda(\tilde{q}_0^2+\nabla^2)^2\right]\phi
%+\frac{u}{4}\phi^4\bigg\},
%\end{equation}
where $a$, $\lambda$, $\tilde{g}$, and $u$ set the magnitudes of the corresponding energetic
contributions, the expressions of which have been derived from either classical DFT
\cite{elder2007phase} or DDFT \cite{vanteeffelen2009derivation,HuangPRE10} in terms of
microscopic direct correlation functions. $\Delta T$ in general is a measure for the temperature
difference from the melting point, while $1/\tilde{q}_0$ is related to the length scale of the
crystalline-type structures that emerge, the properties of which have been studied within the
PFC description. The PFC framework has previously been supported by comparison with DDFT solutions
\cite{vanteeffelen2009derivation} and molecular dynamics simulations \cite{tupper2008phase}.
We remark that in principle the cubic term $\tilde{g}\phi^3/3$ in Eq.~(\ref{eq_Fpfc}) can be
effectively removed by choosing a specific reference point for $\phi$, while we still keep this
term in the present formulation for completeness. (In our calculations in the main text, the
coefficient $\tilde{g}$ is set to zero.)

Second, for the $\tilde{\mathbf{P}}$-dependent part we include
\begin{equation}
\label{eq_FP}
\tilde{\mathcal{F}}_{\tilde{\mathbf{P}}}=
\int\mathrm{d}\mathbf{r}\, \bigg( \frac{1}{2}\tilde{C}_1\left|\bar{\rho}\tilde{\mathbf{P}}\right|^2
  +\frac{1}{4}\tilde{C}_4\left|\bar{\rho}\tilde{\mathbf{P}}\right|^4\bigg), 
\end{equation}
as analyzed in theories of flocking \cite{toner1995long,toner1998flocks}. Again,
$\tilde{C}_1$ and $\tilde{C}_4$ quantify the magnitudes of the corresponding energetic contributions,
for which likewise expressions in terms of microscopic correlation functions are available
\cite{lowen2010phase,wittkowski2010derivation, wittkowski2011polar,wittkowski2011microscopic}.
A combination of $\tilde{C}_1<0$ and $\tilde{C}_4>0$ implies spontaneous polarization supporting
collective motion. It could arise on the microscopic level, for instance, from a pairwise
alignment potential $V^{\mathrm{or}}(\mathbf{\hat{u}}_i,\mathbf{\hat{u}}_j)$
\cite{peruani2010cluster,menzel2012collective}, if in Eq.~(\ref{eq_UVPhi}) we split the pairwise
interaction potential $V$ into a purely positional and a purely orientational part, i.e.,
$V(\mathbf{r}_i,\mathbf{r}_j,\mathbf{\hat{u}}_i,\mathbf{\hat{u}}_j)
= V^{\mathrm{pos}}(\mathbf{r}_i,\mathbf{r}_j) + V^{\mathrm{or}}(\mathbf{\hat{u}}_i,\mathbf{\hat{u}}_j)$.
This type of splitting is reasonable, for example, for spherical particles. In the present work,
we do not consider predominant alignment interactions and instead focus on the regime $\tilde{C}_1>0$.
If only concentrating on the lowest-order contribution in Eq.~(\ref{eq_FP}), we may therefore set
$\tilde{C}_4=0$ as adopted in the simulations described in the main text. The orientational
decorrelation associated with $\tilde{C}_1>0$ results on the microscopic particle level,
for example, from basic rotational diffusion. 

Finally, the expressions in Eqs.~(\ref{eq_Fpfc}) and (\ref{eq_FP}) are inserted via
$\mathcal{F}=\tilde{\mathcal{F}}_{\mathrm{pfc}}+\tilde{\mathcal{F}}_{\tilde{\mathbf{P}}}$ into
Eqs.~(\ref{eq_tildephi1}) and (\ref{eq_tildeP}). Next, the functional derivatives are evaluated,
followed by two sequences of rescaling. In the first step, the rescalings set in Ref.~\cite{elder2004modeling}
are applied, i.e., 
\begin{eqnarray}
  \phi=\phi_0+\bar{\rho}\tilde{\phi}_1&=&
  \sqrt{\lambda\tilde{q}_0^4/u}\,\left(\psi_0+\psi_1\right)=\sqrt{\lambda\tilde{q}_0^4/u}\,\psi,
  \nonumber\\
  a\,\Delta T&=&\lambda\tilde{q}_0^4\epsilon,
\end{eqnarray} 
together with a rescaling of length scale by $1/\tilde{q}_0$. In the second step, 
we apply 
\begin{eqnarray}
v&=&\left(\beta\tilde{D}\bar{\rho}\,\lambda\tilde{q}_0^5/\sqrt{2}\pi\right)\,v_0 \nonumber\\
\tilde{\mathbf{M}}&=&\left(\beta\tilde{D}\bar{\rho}\,\lambda\tilde{q}_0^6/2\pi\right)\,\mathbf{M},
\nonumber\\
\tilde{C}_1&=&\left(\lambda\tilde{q}_0^4/2\right)\,C_1, \nonumber\\
\tilde{C}_4&=&\left(u/4\right)\,C_4, \nonumber\\ %\quad\text{(if included)}, \\
\tilde{g} &=& \sqrt{\lambda\tilde{q}_0^4u}\,g, \nonumber\\
\tilde{D}_r&=&\tilde{D}\tilde{q}_0^2\,D_r, \nonumber\\ 
\tilde{\mathbf{P}}&=&\left(\tilde{q}_0^2\sqrt{2\lambda}/\bar{\rho}\sqrt{u}\right)\,\mathbf{P},
\end{eqnarray} 
together with a rescaling of time by $2\pi/\beta\tilde{D}\bar{\rho}\,\lambda\tilde{q}_0^6$.
As a result, we finally obtain Eqs.~(1) and (2) of the main text governing the rescaled density
variation field $\psi$ and the rescaled polarization field $\mathbf{P}$. We remark that
in fact the inverse length scale $\tilde{q}_0$ is scaled out by the above procedure.
Nevertheless, to be able to refer to it, we maintain it as $q_0=1$ in Eq.~(1) of the main text.

\subsection{Remarks on hydrodynamic interactions}

Recently, we have considered systems of microscopic particles suspended in a surrounding fluid
under low-Reynolds-number conditions. In such a situation, hydrodynamic interactions
\cite{karrila1991microhydrodynamics,dhont1996introduction} mediated between the particles through
the induced hydrodynamic fluid flows make additional contributions to the particle dynamics.
Especially, the thermal noises $\bm{\xi}_{\mathrm{tr},i}$ and $\bm{\xi}_{\mathrm{rot},i}$ become
coupled through the hydrodynamic background as well. Their strengths now depend on the
particle configurations \cite{ermak1978brownian,hennes2014self,pessot2018binary} and need to
be evaluated at each time step, which is a tedious procedure when evaluating the corresponding
Langevin equations. It thus becomes more favorable to target at a statistical equation for
$\mathcal{P}$, similar to the alternative procedure mentioned above, by including an entropic
potential $k_{\mathrm{B}}T\ln \mathcal{P}$ instead of directly working with the stochastic noise
\cite{doi1986theory}. 

When including hydrodynamic interactions, the translational velocity $\mathbf{v}_i$ and angular
velocity $\bm{\omega}_i$ of the $i$th particle now read
\begin{eqnarray}
\label{eq_vi}
\mathbf{v}_i &=& \sum_{j=1}^{N}\left( \bm{\mu}^{\mathrm{tt}}_{ij}\cdot\mathbf{F}_j
+ \bm{\mu}^{\mathrm{tr}}_{ij}\cdot\mathbf{T}_j \right), \\
\label{eq_oi}
\bm{\omega}_i &=& \sum_{j=1}^{N}\left( \bm{\mu}^{\mathrm{rt}}_{ij}\cdot\mathbf{F}_j
+ \bm{\mu}^{\mathrm{rr}}_{ij}\cdot\mathbf{T}_j \right),
\end{eqnarray}
instead of Eqs.~(\ref{eq_mut}) and (\ref{eq_mur}).
%only $\mathbf{v}_i=\mu^{\mathrm{t}}\mathbf{F}_i$ and $\bm{\omega}_i=\mu^{\mathrm{r}}\mathbf{T}_i$. 
Here the hydrodynamic mobility matrices $\bm{\mu}^{\mathrm{tt}}_{ij}$, $\bm{\mu}^{\mathrm{tr}}_{ij}$,
$\bm{\mu}^{\mathrm{rt}}_{ij}$, and $\bm{\mu}^{\mathrm{rr}}_{ij}$ quantify the translation-translation,
translation-rotation, rotation-translation, and rotation-rotation hydrodynamic couplings between
particles $i$ and $j$ \cite{karrila1991microhydrodynamics,dhont1996introduction}. 
In our case, the force $\mathbf{F}_j$ and torque $\mathbf{T}_j$ on the $j$th particle are still
determined by Eqs.~(\ref{eq_F}) and (\ref{eq_T}). 
%\begin{eqnarray}
%\mathbf{F}_j &=& {}-\nabla_{\mathbf{r}_j}\left(U-k_{\mathrm{B}}T\ln P\right) + F^{\mathrm{a}}\mathbf{\hat{u}}_j, \\
%\label{eq_Tj}
%\mathbf{T}_j &=& {}-\mathbf{\hat{u}}_j\times\nabla_{\mathbf{\hat{u}}_j}\left(U-k_{\mathrm{B}}T\ln P\right) + \mathbf{T}^{\mathrm{a}}_j. 
%\end{eqnarray}
%Here, $U=U(\mathbf{r}_1,...,\mathbf{r}_n,\mathbf{\hat{u}}_1,...,\mathbf{\hat{u}}_N)$ sets the total potential interactions between all particles, $F^{\mathrm{a}}$ is the strength of the active self-propulsion force acting on each particle $j$ along its orientation $\mathbf{\hat{u}}_j$, and $\mathbf{T}^{\mathrm{a}}_j$ is the active torque acting on this particle. 
Substituting Eqs.~(\ref{eq_vi}) and (\ref{eq_oi}) into Eq.~(\ref{eq_P}), we then obtain
\begin{widetext}
\begin{eqnarray}
\frac{\partial \mathcal{P}}{\partial t} &=&
\sum_{i=1}^N\sum_{j=1}^N\bigg\{
\nabla_{\mathbf{r}_i}\cdot
\Big( \bm{\mu}^{\mathrm{tt}}_{ij}\cdot\left[\nabla_{\mathbf{r}_j}
  \left(U+k_{\mathrm{B}}T\ln \mathcal{P}\right) - F^{\mathrm{a}}\mathbf{\hat{u}}_j\right] 
%\nonumber\\
%&&
%\qquad
+ \bm{\mu}^{\mathrm{tr}}_{ij}\cdot\left[\mathbf{\hat{u}}_j\times\nabla_{\mathbf{\hat{u}}_j}
  \left(U+k_{\mathrm{B}}T\ln \mathcal{P}\right) - \mathbf{T}^{\mathrm{a}}_j\right] \Big)\mathcal{P}
\nonumber\\
&&
\quad\qquad{}+(\mathbf{\hat{u}}_j\times\nabla_{\mathbf{\hat{u}}_j})\cdot
\Big( \bm{\mu}^{\mathrm{rt}}_{ij}\cdot\left[\nabla_{\mathbf{r}_j}\left(U+k_{\mathrm{B}}T\ln \mathcal{P}\right)
  - F^{\mathrm{a}}\mathbf{\hat{u}}_j\right] 
%\nonumber\\
%&&
%\qquad
+ \bm{\mu}^{\mathrm{rr}}_{ij}\cdot\left[\mathbf{\hat{u}}_j\times\nabla_{\mathbf{\hat{u}}_j}
  \left(U+k_{\mathrm{B}}T\ln \mathcal{P}\right) - \mathbf{T}^{\mathrm{a}}_j\right] \Big)\mathcal{P}
\bigg\},~\;\quad
\label{eq_Phydro}
\end{eqnarray}
\end{widetext}
which includes the hydrodynamic interactions via the mobility matrices. 

In the bulk fluid, denoting the distance between particles $i$ and $j$ as
$r_{ij}=|\mathbf{r}_i-\mathbf{r}_j|$, the hydrodynamic mobility matrices in the far-field
approximation scale as $\bm{\mu}^{\mathrm{tt}}_{i=j},\bm{\mu}^{\mathrm{rr}}_{i=j}\sim (1/r_{ij})^0=1$,
$\bm{\mu}^{\mathrm{tt}}_{i\neq j}\sim 1/r_{ij}$, $\bm{\mu}^{\mathrm{rt}}_{ij},
\bm{\mu}^{\mathrm{tr}}_{ij}\sim (1/r_{ij})^2$, and $\bm{\mu}^{\mathrm{rr}}_{i\neq j}\sim (1/r_{ij})^3$
for spherical particles \cite{karrila1991microhydrodynamics,dhont1996introduction}. Thus,
inserting $\bm{\mu}^{\mathrm{tt}}_{i=j}=\mu^{\mathrm{t}}\bm{1}$ and $\bm{\mu}^{\mathrm{rr}}_{i=j}
=\mu^{\mathrm{r}}\bm{1}$ \cite{karrila1991microhydrodynamics,dhont1996introduction}, with $\bm{1}$
representing the unit matrix, and focusing only on the zeroth-order effects, Eq.~(\ref{eq_Phydro})
is reduced to Eq.~(\ref{eq_Pdyn}) and the above results for the active PFC description analyzed
in the main text can be recovered. In principle, such a situation could be realized experimentally
in a bulk fluid, for instance, by tracking the positions of colloidal particles in real time
\cite{bauerle2018self,lavergne2019group} and driving accordingly the particles by optical
tweezers \cite{hanes2009combined,williams2016transmission}. In such a setup, the deterministic
trajectories resulting from both self-propulsion and self-spinning as well as from interparticle
interactions would need to be programmed into the optical gearing of the colloidal particles,
plus the effect of positional and orientational noises. %Particle rotations could be superimposed
%using magnetic Janus particles under rotating external magnetic fields \cite{}. 
Instead, in this work we mainly focus on the study of self-propelling and self-spinning active
particles confined on a 2D substrate. Generally, if such a system is embedded in a surrounding
fluid, the substrate can significantly limit the influence of the above-mentioned hydrodynamic
interactions, %(with the leading-order correction of $\mathcal{O}(1/r_{ij})$), 
particularly for hydrodynamic no-slip surfaces. It is thus expected that the results presented
in the main text will in many cases still apply in a fluid environment.

\bibliography{references_suppl_v2}